\documentclass[aps,11pt, superscriptaddress, showpacs,showkeys,floatfix]{revtex4}
\usepackage{graphicx}
\usepackage{colordvi}

\begin{document}

\title{Tunneling probability for the birth of an asymptotically 
DeSitter universe.}

\author{J. Acacio de Barros\footnote{On leave from Departamento de 
Fisica, Universidade Federal de Juiz de Fora, Brazil} 
\footnote{E-mail: jacaciodebarros@gmail.com}}

\address{CSLI, \\ 
220 Panama Street, \\
Stanford University, \\
Stanford, CA 94305-4115, USA.}

\author{E. V. Corr\^{e}a Silva\footnote{E-mail: evasquez@uerj.br}}

\author{G. A. Monerat\footnote{E-mail: monerat@uerj.br}}
 
\author{G. Oliveira-Neto\footnote{E-mail: gilneto@fat.uerj.br}}

\address{Departamento de Matem\'{a}tica e Computa\c{c}\~{a}o, 
Faculdade de Tecnologia, \\ 
Universidade do Estado do Rio de Janeiro, Estrada Resende-Riachuelo, 
s/n$^o$, Morada da Colina \\
CEP 27523-000, Resende-RJ, Brazil.}

\author{L. G. Ferreira Filho\footnote{E-mail: gonzaga@fat.uerj.br}}

\address{Departamento de Mec\^{a}nica e Energia, 
Faculdade de Tecnologia, \\ 
Universidade do Estado do Rio de Janeiro, Estrada Resende-Riachuelo, 
s/n$^o$, Morada da Colina \\
CEP 27523-000, Resende-RJ, Brazil.}

\author{P. Romildo Jr.\footnote{E-mail: pauloromildo@yahoo.com.br}}

\address{Departamento de F\'{\i}sica, Instituto de Ci\^{e}ncias Exatas, 
Universidade Federal de Juiz de Fora, \\
CEP 36036-330, Juiz de Fora, Minas Gerais, Brazil.}

\date{\today}

\begin{abstract}
In the present work, we quantize a closed Friedmann-Robertson-Walker model 
in the presence of a positive cosmological constant and radiation. It gives 
rise to a Wheeler-DeWitt equation for the scale factor which has the form 
of a Schr\"{o}dinger equation for a potential with a barrier. We solve it
numerically and determine the tunneling probability for the birth of a 
asymptotically DeSitter, inflationary universe, initially, as a function 
of the mean energy of the initial wave-function. Then, we verify that the 
tunneling probability increases with the cosmological constant, for a fixed 
value of the mean energy of the initial wave-function. Our treatment of the 
problem is more general than previous ones, based on the WKB approximation.
That is the case because we take into account the fact that the scale factor 
($a$) cannot be smaller than zero. It means that, one has to introduce an 
infinity potential wall at $a = 0$, which forces any wave-packet to be zero 
there. That condition introduces new results, in comparison with previous
works.
\end{abstract}



\maketitle

\section{Introduction}
\label{sec:intro}

Since the pioneering work in quantum cosmology due to DeWitt \cite{dewitt},
many physicists have worked in this theory. The main motivation behind 
quantum cosmology is a consistent explanation for the origin of our Universe.
So far, the most appealing explanation is the spontaneous {\it creation from 
nothing} \cite{grishchuk, vilenkin, hawking, linde, rubakov, vilenkin1}.
In that picture for the origin of the Universe, the Universe is a quantum
mechanical system with zero size. There is a potential barrier that the 
Universe may tunnel with a well-defined, non-zero probability. If the
Universe actually tunnels, it emerges to the right of the barrier with a
definite size. The application of the {\it creation from nothing} idea in 
minisuperspace models has led to several important results. The wave-function 
of the Universe satisfies the Wheeler-DeWitt equation \cite{wheeler, dewitt}. 
Therefore, one needs to specify boundary conditions in order to solve that 
equation and find a unique and well-defined wave function. The motivation to 
obtain a wave-function that represents the {\it creation from nothing} has led 
to the introduction of at least three proposals for the boundary conditions for 
the wave-function of the Universe \cite{vilenkin1}. The inflationary period of 
the Universe appears very naturally from the {\it creation from nothing} idea. 
That is the case because most of the minisuperspace models considered so far
have a potential that decreases, without a limit, to the right of the barrier.
It gives rise to a period of unbounded expansion which is interpreted as the 
inflationary period of the Universe \cite{vilenkin1}. Also, it was shown by
several authors that an open inflationary universe may be created from nothing,
in theories of a single scalar field for generic potentials \cite{hawking1,
linde1, bousso}. Another important issue is the particle content in the
Universe originated during the {\it creation from nothing} process \cite{rubakov,
rubakov1, vilenkin2}.

In the present work, we would like to explicitly compute the quantum mechanical 
probability that the universe tunnels through a potential barrier and initiates
its classical evolution. That probability is the tunneling probability (TP) and the
particular model we consider, here, is a closed Friedmann-Robertson-Walker (FRW) 
model in the presence of a positive cosmological constant and radiation. The 
radiation is treated by means of the variational formalism developed by Schutz 
\cite{schutz}. That model has already been treated quantum mechanically using the 
ADM formalism and the Dirac quantization for constrained systems 
\cite{vilenkin1, bertolami, cavaglia}. The wave-function, for that model, was 
calculated in the WKB approximation. Here, we compute the wave-function and TP 
exactly, without any approximation. It will be done by means of a numerical 
calculation. In particular, our treatment of the problem is more general than 
previous ones, because we take into account the fact that the scale factor ($a$) 
cannot be smaller than zero. It means that, one has to introduce an infinity 
potential wall at $a = 0$, which forces any wave-packet to be zero there. As we 
shall see, that condition introduces new results, in comparison with previous
works. This model has two free parameters: the radiation energy and the 
cosmological constant. Therefore, we will obtain the TP as a function of those 
two parameters. One of the main motivations of any quantum cosmological model 
is to fix initial conditions for the classical evolution of our Universe 
\cite{grishchuk}. Here, for the present model, we would gain information on
what is the most probable amount of radiation in the initial evolution of the 
classical universe and the most probable value of the cosmological constant.
Another motivation of the present work, is trying to contribute to a long 
standing debate about which is the most appropriate set of initial conditions for 
the wave-function of the Universe. The most well-known proposals for the set of 
initial conditions are the {\it tunneling} one, due to A. Vilenkin 
\cite{vilenkin}, and the {\it no-boundary} one, due to  J. B. Hartle and S. W. 
Hawking \cite{hawking}. The application of those proposals for simple models showed 
that they give some different predictions for the initial evolution of the Universe
\cite{vilenkin, hawking, vilenkin2, bertolami, cavaglia, gil, moss}. One of such 
predictions, which we shall explore here, is the initial energy of the Universe 
right after its nucleation. The {\it tunneling} wave function predicts that the 
Universe must nucleate with the largest possible vacuum energy whereas the 
{\it no-boundary} wave function predicts just the opposite \cite{vilenkin1}. In
terms of our results, if one assumes that the cosmological constant describes a 
vacuum energy, it is important to see if TP increases or decreases with the 
cosmological constant, for fixed radiation energy. The first behavior favors the 
{\it tunneling} wave function and the latter favors the {\it no-boundary} wave 
function.

In the next Section, we describe the classical dynamics of the present cosmological
model. We write the super-hamiltonian constraint and the Hamilton's equations. We
solve the Hamilton's equations and find the general solution of the model. Then, we
qualitatively describe all possible classical evolutions. In Section 
\ref{sec:quantum}, we canonically quantize the model and obtain the corresponding
Wheeler-DeWitt equation. We solve it, numerically, for particular values of the
radiation energy and the cosmological constant. We show the square modulus of the
wave-function of the universe as a function of the scalar factor. The tunneling
process can be readily seen from that figure. The Section \ref{sec:tp} is divided
in three subsections with the main results of the paper. In the first subsection
\ref{subsec:tpenergy}, we start introducing the tunneling probability, then we 
evaluate its dependence on the radiation energy. We obtain that, the TP increases 
with the radiation energy for a fixed cosmological constant. Therefore, it is
more probable that the classical evolution should start with the greatest possible
value for the radiation energy. In the following subsection \ref{subsec:wkb}, we 
give a detailed comparison between the exact TP, computed in the previous subsection, 
and the corresponding WKB tunneling probability. Here, we show how the presence of
an infinity potential wall at $a=0$ may lead to a difference between our results and 
previous ones, based on the $WKB$ approximation. In the final subsection 
\ref{subsec:tplambda} of this section, we evaluate the dependence of the exact TP with 
the cosmological constant. We obtain that, the TP increases with the cosmological 
constant for a fixed radiation energy. Therefore, it is more probable that the 
classical evolution should start with the greatest possible value for the cosmological
constant. This behavior of $TP$ also favors the {\it tunneling} wave function. Finally, 
in Section \ref{sec:conclusions} we summarize the main points and results of our paper.

\section{The Classical Model}
\label{sec:classical}

The Friedmann-Robertson-Walker cosmological models are characterized by the
scale factor $a(t)$ and have the following line element,

\begin{equation}  \label{1}
ds^2 = - N^2(t) dt^2 + a^2(t)\left( \frac{dr^2}{1 - kr^2} + r^2 d\Omega^2
\right)\, ,
\end{equation}
where $d\Omega^2$ is the line element of the two-dimensional sphere with
unitary radius, $N(t)$ is the lapse function and $k$ gives the type of
constant curvature of the spatial sections. Here, we are considering the 
case with positive curvature $k=1$ and we are using the natural
unit system, where $\hbar=c=G=1$. The matter content of the model is
represented by a perfect fluid with four-velocity $U^\mu = \delta^{\mu}_0$
in the comoving coordinate system used, plus a positive cosmological
constant. The total energy-momentum tensor is given by,

\begin{equation}
T_{\mu,\, \nu} = (\rho+p)U_{\mu}U_{\nu} - p g_{\mu,\, \nu} - \Lambda
g_{\mu,\, \nu}\, ,  \label{2}
\end{equation}
where $\rho$ and $p$ are the energy density and pressure of the fluid,
respectively. Here, we assume that $p = \rho/3$, which is the equation of
state for radiation. This choice may be considered as a first approximation
to treat the matter content of the early Universe  and it was made as a
matter of simplicity. It is clear that a more complete treatment should
describe the radiation, present in the primordial Universe, in terms of the
electromagnetic field.

Einstein's equations for the metric (\ref{1}) and the energy momentum tensor
(\ref{2}) are equivalent to the Hamilton equations generated by the
super-hamiltonian constraint

\begin{equation}
{\mathcal{H}}= -\frac{p_{a}^2}{12 a} - 3a +\Lambda a^{3} + {p_{T}\over a},  
\label{3}
\end{equation}
where $p_{a}$ and $p_{T}$ are the momenta canonically conjugated to $a$ and 
$T$ the latter being the canonical variable associated to the fluid \cite
{germano1}. The total Hamiltonian is given by $N {\mathcal{H}}$ and we shall
work in the conformal gauge, where $N = a$.

The classical dynamics is governed by the Hamilton equations, derived from
the total Hamiltonian $N {\mathcal{H}}$, namely

\begin{equation}
\left\{ 
\begin{array}{llllll}
\dot{a} = & \frac{\partial (\displaystyle N\mathcal{H})}{\displaystyle 
\partial p_{a}}=-\frac{\displaystyle p_{a}}{\displaystyle 6}\, , &  &  &  & 
\\ 
&  &  &  &  &  \\ 
\dot{p_{a}} = & -\frac{\displaystyle \partial (N\mathcal{H})}{\displaystyle 
\partial a}=6a - 4\Lambda a^3 \, , &  &  &  &  \\ 
&  &  &  &  &  \\ 
\dot{T} = & \frac{\displaystyle \partial (N\mathcal{H})}{\displaystyle 
\partial p_{T}} = 1\, , &  &  &  &  \\ 
&  &  &  &  &  \\ 
\dot{p_{T}} = & -\frac{\displaystyle \partial (N\mathcal{H})}{\displaystyle 
\partial T}=0\, . &  &  &  &  \\ 
&  &  &  &  & 
\end{array}
\right.  \label{4}
\end{equation}
Where the dot means derivative with respect to the conformal time $\tau \equiv
Nt$. We also have the constraint equation $\mathcal{H} = 0$. We have the 
following solutions for the system (\ref{4}):

\begin{equation}
T (\tau) = \tau + c_{1}\, , \\
a(\tau) = \frac{\sqrt{6\,\beta}}{{{\displaystyle\sqrt{3+\sqrt
{9-12\,\Lambda\, \beta}}}}} sn\left(\frac{\displaystyle\sqrt{18+6\,\sqrt{
9-12\, \Lambda\,\beta}} \left( \tau-\tau_0\right)}{\displaystyle 6}
,\sigma\right),  \label{5}
\end{equation}
where $c_{1}$, $\beta$ and $\tau_{0}$ are integration constants, $sn$ is the
Jacobi's elliptic sine \cite{abramowitz} of modulus $\sigma$ given by

\begin{equation}
\sigma =\frac{\sqrt{2}}{2}\sqrt{{\frac{-2\beta\Lambda+3-\sqrt{9-12\,
\Lambda\,\beta}}{\Lambda\,\beta}}}.  \label{6}
\end{equation}
In the case studied here $\Lambda > 0$, the radiation energy $\beta$
can take values in the domain, $\beta \leq 3/(4\Lambda)$. If one 
substitutes values of $\beta$ such that $\beta < 3/(4\Lambda)$ in 
Eqs. (\ref{5}) and (\ref{6}), the scale factor, starting from zero, 
expands to a maximum size and then recollapse. On the other hand, if 
$\beta = 3/(4\Lambda)$, the scalar factor initially decelerates and 
then, enter the regime of unbounded expansion.

\section{The Quantum Model}
\label{sec:quantum}

We wish to quantize the model following the Dirac formalism for quantizing
constrained systems \cite{dirac}. First we introduce a wave-function which
is a function of the canonical variables $\hat{a}$ and $\hat{T}$,

\begin{equation}  \label{7}
\Psi\, =\, \Psi(\hat{a} ,\hat{T} )\, .
\end{equation}
Then, we impose the appropriate commutators between the operators $\hat{a}$
and $\hat{T}$ and their conjugate momenta $\hat{P}_a$ and $\hat{P}_T$.
Working in the Schr\"{o}dinger picture, the operators $\hat{a}$ and $\hat{T}$
are simply multiplication operators, while their conjugate momenta are
represented by the differential operators 
\begin{equation}
p_{a}\rightarrow -i\frac{\partial}{\partial a}\hspace{0.2cm},\hspace{0.2cm} 
\hspace{0.2cm}p_{T}\rightarrow -i\frac{\partial}{\partial T}\hspace{0.2cm}.
\label{8}
\end{equation}

Finally, we demand that the operator corresponding to $N \mathcal{H}$ 
annihilate the wave-function $\Psi$, which leads to Wheeler-DeWitt 
equation 
\begin{equation}
\bigg(\frac{1}{12}\frac{{\partial}^2}{\partial a^2} - 3a^2 + \Lambda a^4
\bigg)\Psi(a,\tau) = -i \, \frac{\partial}{\partial \tau}\Psi(a,\tau),
\label{9}
\end{equation}
where the new variable $\tau= -T$ has been introduced.

The operator $N \hat{\mathcal{H}}$ is self-adjoint \cite{lemos} with respect
to the internal product,

\begin{equation}
(\Psi ,\Phi ) = \int_0^{\infty} da\, \,\Psi(a,\tau)^*\, \Phi (a,\tau)\, ,
\label{10}
\end{equation}
if the wave functions are restricted to the set of those satisfying either 
$\Psi (0,\tau )=0$ or $\Psi^{\prime}(0, \tau)=0$, where the prime $\prime$
means the partial derivative with respect to $a$. Here, we consider wave 
functions satisfying the former type of boundary condition and we also 
demand that they vanish when $a$ goes to $\infty$.

The Wheeler-DeWitt equation (\ref{9}) is a Schr\"{o}dinger equation for
a potential with a barrier. We solve it numerically using a finite
difference procedure based on the Crank-Nicholson method \cite{crank}, 
\cite{numericalrecipes} and implemented in the program GNU-OCTAVE. Our
choice of the Crank-Nicholson method is based on its recognized stability.
The norm conservation is commonly used as a criterion to evaluate the
reliability of the numerical calculations of the time evolution of
wave functions. In References \cite{Iitaka} and \cite{teukolsky}, this 
criterion is used to show analytically that the Crank-Nicholson method is 
unconditionally stable. Here, in order to evaluate the reliability of our 
algorithm, we have numerically calculated the norm of the wave packet for 
different times. The results thus obtained show that the norm is preserved.
  
In fact, numerically one can only treat
the {\it tunneling from something} process, where one gives a initial wave
function with a given mean energy, very concentrated in a region next to 
$a=0$. That initial condition fixes an energy for the radiation and the 
initial region where $a$ may take values. Our choice for the initial wave
function is the following normalized gaussian,

\begin{equation}
\label{11}
\Psi(a,0) = \left({8192 E^3\over \pi}\right)^{1/4} a 
e^{(-4 E a^2)}\, ,
\end{equation}
where $E$ is the radiation energy. $\Psi(a,0)$ is normalized by demanding 
that the integral of $|\Psi(a,0)|^2$ from $0$ to $\infty$ be equal to one
and its mean energy be $E$. After one gives the initial wave function,
one leaves it evolve following the appropriate Schr\"{o}dinger equation
until it reaches infinity in the $a$ direction. Numerically, one has to 
fix the infinity at a finite value. In the present case we fix $a_{max}=30$
as the infinity in the $a$ direction. The general behavior of the solutions,
when E is smaller than the maximum value of the potential barrier, is an
everywhere well-defined, finite, normalized wave packet. Even in the limit
when the scale factor goes to zero. For small values of $a$ the wave packet
have great amplitudes and oscillates rapidly due to the interaction between
the incident and reflected components. The transmitted component is an
oscillatory wave packet that moves to the right and has a decreasing 
amplitude which goes to zero in the limit when $a$ goes to infinity.
As an example, we solve eq. ({\ref{9}) with $\Lambda=0.0121$. For this
choice of $\Lambda$ the potential barrier has its maximum value equal to
185.95. In order to see the tunneling process, we choose $E=185$ for the
initial wave function eq. (\ref{11}). For that energy, we compute the points
where it meets the potential barrier, the left ($a_{ltp}$) and right
($a_{rtp}$) turning points. They are, $a_{ltp}=10.7287$ and $a_{rtp}=11.5252$. 
In figure \ref{fig1}, we show $|\Psi(a,t_{max})|^2$ for the values of
$\Lambda$ and $E$, given above, at the moment ($t_{max}$) when $\Psi$ reaches
infinity. For more data on this particular case see table \ref{tablelambda},
in the appendix. It is important to mention that the choice of the numerical 
values for $\Lambda$ and $E$ above and in the following examples, in the next 
section, were made simply for a better visualization of the different 
properties of the system.

\begin{figure}[h!]
\includegraphics[width=7cm,height=9cm,angle=-90]{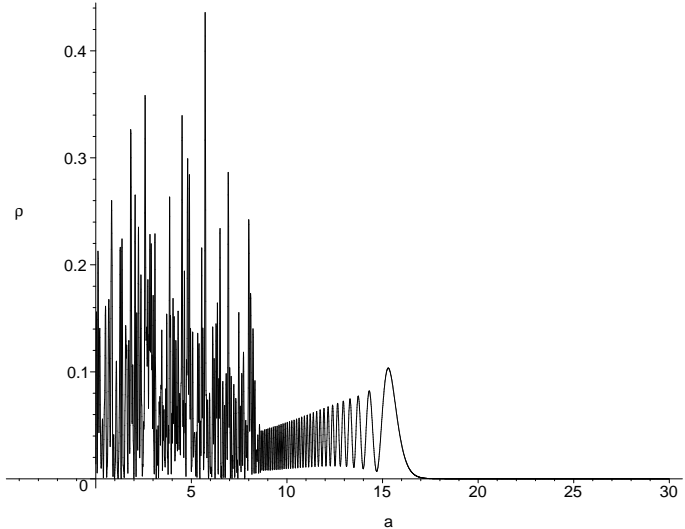}
\caption{{\protect\footnotesize {$|\Psi(a,t_{max})|^2 \equiv \rho$, for 
$\Lambda=0.0121$, $E=185$ at the moment $t_{max}$ when $\Psi$ reaches 
infinity, located at $a=30$.}}}
\label{fig1}
\end{figure}

\section{Tunneling Probabilities}
\label{sec:tp}

\subsection{Tunneling probability as a function of $E$}
\label{subsec:tpenergy}

We compute the tunneling probability as the probability to find the 
scale factor of the universe to the right of the potential barrier.
In the present situation, this definition is given by the following 
expression,

\begin{equation}
\label{12}
TP_{int} = {\int_{a_{rtp}}^{\infty} |\Psi(a,tmax)|^2 da \over 
\int_{0}^{\infty} |\Psi(a,tmax)|^2 da} \, ,
\end{equation}
where, as we have mentioned above, numerically $\infty$ has to be fixed to
a maximum value of $a$. Here, we are working with $a_{max} = 30$.

Since, by normalization, the denominator of Eq. (\ref{12}) is equal to the
identity, $TP_{int}$ is effectively given by the numerator of Eq. (\ref{12}). 
We consider initially the dependence of TP on the energy $E$. Therefore, we 
compute $TP_{int}$ for many different values of $E$ for a fixed $\Lambda$.
For all cases, we consider the situation where $E$ is smaller than the
maximum value of the potential barrier. From that numerical study we 
conclude that the tunneling probability grows with $E$ for a fixed $\Lambda$.
As an example, we consider $47$ values of the radiation energy for a fixed 
$\Lambda=0.01$. For this choice of $\Lambda$ the potential barrier has its 
maximum value equal to 225. In order to study the tunneling process, we fix 
the mean energies of the various $\Psi(a,0)$'s eq. (\ref{11}) to be smaller 
than that value. In table \ref{tableenergy}, in the appendix, we can see, 
among other quantities, the different values of the energy $E$,  $TP_{int}$, 
$a_{ltp}$ and $a_{rtp}$ for each energy. In figure \ref{fig2}, we see the 
tunneling probability as functions of $E$, for this particular example. Due 
to the small values of some $TP'$s, we plot the logarithms of the $TP'$s 
against $E$. 

\begin{figure}[h!]
\includegraphics[width=9cm,height=11cm,angle=-90]{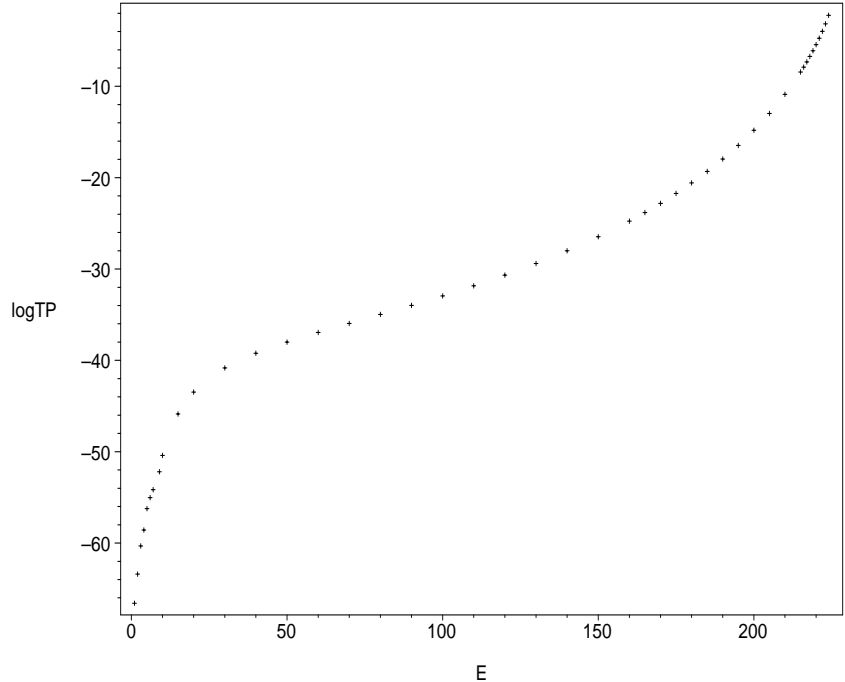}
\caption{{\protect\footnotesize {
$\log TP_{int}$  
for different radiation energies ($E$) for a fixed $\Lambda=0.01$.}}}
\label{fig2}
\end{figure}

Since $TP$ grows with $E$ it is more likely for the universe, described by 
the present model, to nucleate with the highest possible radiation energy.
Therefore, it is more probable that the classical evolution should start 
with the greatest possible value for the radiation energy.

\subsection{Exact tunneling probability versus WKB tunneling probability}
\label{subsec:wkb}

Let us, now, compare the exact tunneling probability represented by
$TP_{int}$ eq. (\ref{12}) with the approximated WKB tunneling probability
($TP_{WKB}$). The $TP_{WKB}$ is defined by the ratio between the square 
modulus of the transmitted amplitude of the WKB wave-function and the 
square modulus of the incident amplitude of the WKB wave-function 
\cite{griffths}, \cite{merzbacher}. For the present situation, it is 
given by the following expression \cite{merzbacher},

\begin{equation}
\label{13}
TP_{WKB} = {4\over \left(2\theta+{1\over(2\theta)}\right)^2}\, ,
\end{equation}
where,
\begin{equation}
\label{14}
\theta = \exp{\left(\int_{a_{ltp}}^{a_{rtp}} da \sqrt{12(3a^2-\Lambda a^4 
- E)}\right)}\, .
\end{equation}

It is important to note that the $TP_{WKB}$ eq. (\ref{13}), was computed
considering that the incident wave ($\Psi_I$) reaches the potential barrier 
at $a_{ltp}$. Then, part of $\Psi_I$ is transmitted to $\infty$ ($\Psi_T$)
and part is reflected to $-\infty$ ($\Psi_R$). In the present problem, we 
have an infinity potential wall at $a=0$ because the scale factor 
cannot be smaller than zero. It means that $\Psi_R$ cannot go 
to $-\infty$, as was assumed in order to compute the $TP_{WKB}$. Instead, 
$\Psi_R$ will reach the infinity potential wall at $a=0$ and will be entirely
reflected back toward the potential barrier, giving rise to a new incident
wave ($(\Psi_R)_I$). The new incident wave $(\Psi_R)_I$ reaches the
potential barrier at $a_{ltp}$ and is divided in two components. A
reflected component which moves toward the infinity potential wall at
$a=0$ ($((\Psi_R)_I)_R$) and a transmitted component which moves toward
$\infty$ ($((\Psi_R)_I)_T$). $((\Psi_R)_I)_T$ will contribute a new amount
to the already existing $TP_{int}$ due to ($\Psi_T$). On the other hand, 
$((\Psi_R)_I)_R$, after being reflected at $a=0$, gives rise to a new incident
wave which will contribute a further amount to the already existing $TP_{int}$.
If we let our system evolve for a long period of time, $TP_{int}$ will get 
many such contributions from the different reflected components. The only way
it makes sense comparing $TP_{int}$ with $TP_{WKB}$ is when we let the system
evolve for a period of time ($\Delta t$) during which $\Psi_R$ cannot be 
reflected at $a=0$ and come back to reach the potential barrier. It is clear
by the shape of our potential that the greater the mean energy $E$ of the
wave-packet (\ref{11}), the greater is ($\Delta t$). As an example, in Table 
\ref{comparison}, we show a comparison between $TP_{int}$ and $TP_{WKB}$ for 
different values of $E$ and $\Delta t$ for the case with $\Lambda = 0.01$. We 
can see, clearly, that both tunneling probabilities coincide if we consider 
the appropriate $\Delta t$, for each $E$.

\begin{table}[h!]
{\scriptsize\begin{tabular}{|c|c|c|c|}
\hline $E$ & $TP_{int}$ & $TP_{WKB}$ & $\Delta t$ \\ \hline
$80$ & $6.0648\times 10^{-302}$ & $1.1845\times 10^{-303}$ & $13$\\ \hline
$100$ & $1.5887\times 10^{-259}$ & $3.7271\times 10^{-258}$ & $18.5$\\ \hline
$130$ & $6.4375\times 10^{-194}$ & $9.8051\times 10^{-193}$ & $30$\\ \hline
$160$ & $6.9194\times 10^{-130}$ & $5.7774\times 10^{-130}$ & $45.5$\\ \hline
$175$ & $6.5061\times 10^{-100}$ & $2.1361\times 10^{-99}$ & $54.5$\\ \hline
$190$ & $5.3119\times 10^{-69}$ & $2.6372\times 10^{-69}$ & $65.5$\\ \hline
$200$ & $2.2682\times 10^{-49}$ & $1.7295\times 10^{-49}$ & $73.5$\\ \hline
$215$ & $5.4531\times 10^{-20}$ & $4.1983\times 10^{-20}$ & $88$\\ \hline
$219$ & $5.2168\times 10^{-12}$ & $2.4754\times 10^{-12}$ & $93$\\ \hline
$223$ & $7.0045\times 10^{-04}$ & $1.3731\times 10^{-04}$ & $100$\\ \hline
\end{tabular}
}
\caption{{\protect\footnotesize {A comparison between $TP_{int}$ and $TP_{WKB}$
for $10$ different values of $E$ with its associated integration time $\Delta t$ 
for the case with $\Lambda = 0.01$.
}}}
\label{comparison}
\end{table}

In order to have an idea on how the $TP_{int}$ may differ from the $TP_{WKB}$,
we let the initial wave-packet (\ref{11}), with different mean energies, evolve 
during the same time interval $\Delta t$. We consider the example given in the 
previous subsection \ref{subsec:tpenergy}, with a common time interval of 100.
Observing Table \ref{comparison}, we see that this amount of $\Delta t$ 
guarantees that $\Psi_R$ of the wave-packet with mean energy $223$ does not
contribute to the $TP_{int}$. Therefore, we may expect that to all wave-packets
with mean energies smaller than $223$, $TP_{int}$ will be greater than $TP_{WKB}$.
We show this comparison in table \ref{tableenergy}, in the appendix, where we have 
an entry for $TP_{WKB}$. It means that, we computed the values of the $TP_{WKB}$s 
for each energy used to compute the $TP_{int}$s, in the case where $\Lambda = 0.01$. 
In figure \ref{fig3}, we show, graphically, that comparison between both tunneling 
probabilities as functions of $E$, with $\Delta t = 100$ for all values of $E$. Due 
to the small values of some $TP'$s, we plot the logarithms of the $TP'$s against $E$. 

\begin{figure}[h!]
\includegraphics[width=9cm,height=11cm,angle=-90]{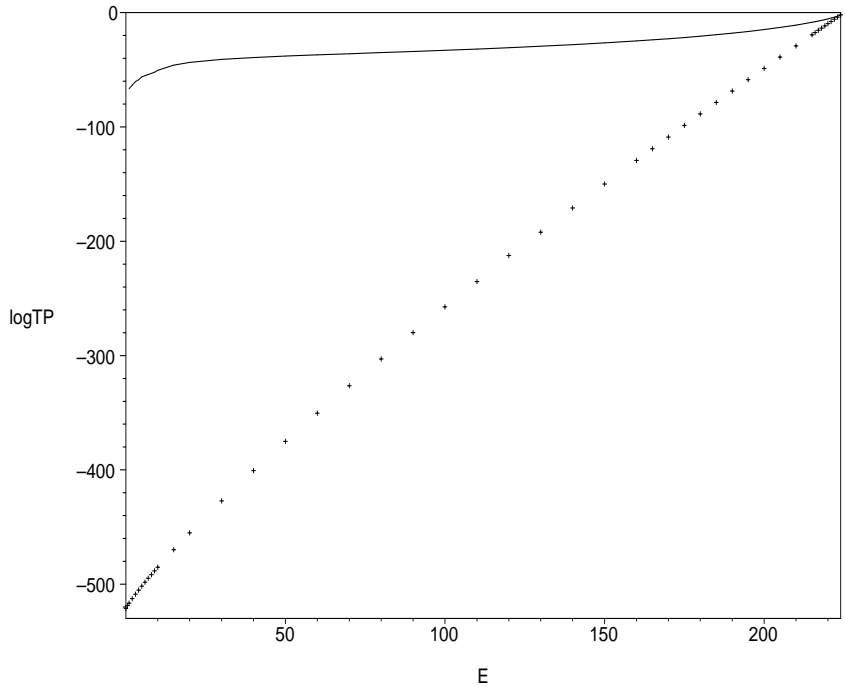}
\caption{{\protect\footnotesize {Comparison between $\log TP_{WKB}$ (dots) 
and $\log TP_{int}$ (line) for different radiation energies ($E$) for a 
fixed $\Lambda=0.01$. The integration time $\Delta t$ is equal to $100$ for all 
values of $E$.}}}
\label{fig3}
\end{figure}

As we can see from figure \ref{fig3}, for this choice of $\Delta t$ the 
tunneling probabilities disagree for most values of $E$. They only agree for
values of $E$ very close to the top of the potential barrier. There, because
the values of $E$ are similar to $223$, $\Delta t$ is almost sufficient to 
guarantee that $\Psi_R$ of each wave-packet does not contribute to the 
$TP_{int}$.

As we have mentioned above, numerically one can only treat the {\it 
tunneling from something} process, where one gives a initial wave function 
with a given mean energy, very concentrated in a region next to $a=0$. 
Then, if we take $E=0$ the $TP_{int}$ will be zero. On the other hand,
we may have an idea how $TP_{int}$ behaves near $E=0$ from figure \ref{fig2}.
From table \ref{tableenergy}, in the appendix,
one can see that $TP_{WKB}=7.0246\times 10^{-522}$ when $E=0$.

Finally, we may compute the time ($\tau$) the universe would take, for
each energy, to nucleate. In order to understand the meaning of $\tau$, 
consider a photon that composes the radiation which is initially confined 
to the left of the potential barrier. Then, compute the emission probability 
of that photon, per unit of time. We may invert it to obtain $\tau$, the 
time the photon would take to escape the potential barrier. If we consider 
$\tau$ as the time it takes for the most part of the photons to escape the 
barrier, we obtain the time the universe would take, for each energy, to 
appear at the right of the barrier. In the present situation, $\tau$ is 
given by the following expression \cite{griffths},

\begin{equation}
\label{15}
\tau = 2 a_{ltp} {1\over PT_{int}}
\end{equation}

From table \ref{tableenergy}, in the appendix, we may see the 
values of $\tau$, for each energy. It is clear by the results that, the 
smaller the energy $E$ the longer it will take for the universe to 
nucleate.

\subsection{Tunneling probability as a function of $\Lambda$}
\label{subsec:tplambda}

We would like to study, now, how the tunneling probability depends on
the cosmological constant. In order to do that, we must fix an initial
energy $E$ for the radiation and compute the $TP_{int}$ for various 
values of $\Lambda$. For all cases, we consider the situation where $E$ 
is smaller than the maximum value of the potential barrier. From that 
numerical study we conclude that the tunneling probability grows with 
$\Lambda$ for a fixed $E$. As an example, we choose $E=185$ and $22$ 
different values of $\Lambda$, such that, the maximum energy of the 
potential barrier ($PE_{max}$), for each $\Lambda$, is greater than 
$185$. The values of $\Lambda$, $TP_{int}$, $PE_{max}$, $\tau$, $a_{ltp}$ 
and $a_{rtp}$ are given in table \ref{tablelambda}, in the appendix. 
With those values, we construct the curve $TP_{int}$ versus $\Lambda$, 
shown in figure \ref{fig6}. Due to the small values of some $TP'$s, we 
plot the logarithms of the $TP'$s against $\Lambda$.

Since $TP$ grows with $\Lambda$ it is more likely for the universe, 
described by the present model, to nucleate with the highest possible
cosmological constant. Therefore, it is more probable that the classical 
evolution should start with the greatest possible value for the cosmological
constant. Also, if we assume that $\Lambda$ describes a vacuum energy, this 
result is qualitatively in accordance with the prediction of the 
{\it tunneling} wave function due to A. Vilenkin \cite{vilenkin}.

\begin{figure}[h!]
\includegraphics[width=12cm,height=10cm,angle=0]{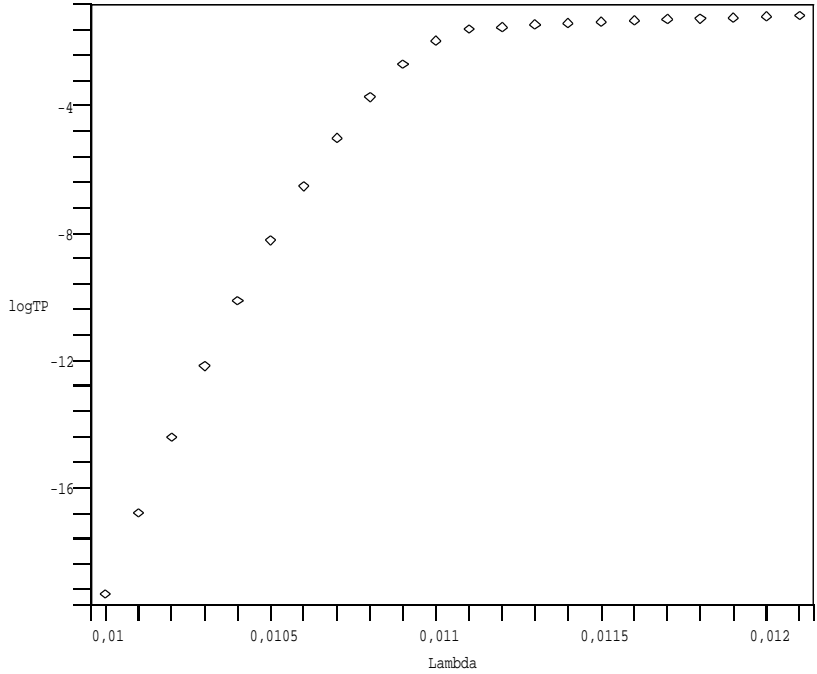}
\caption{{\protect\footnotesize {$\log TP_{int}$ for $22$ different values 
of $\Lambda$ for a fixed $E=185$.}}}
\label{fig6}
\end{figure}

\section{Conclusions.}
\label{sec:conclusions}

In the present work, we canonically quantized  
a closed Friedmann-Robertson-Walker (FRW) model in the presence of a positive 
cosmological constant and radiation. The radiation was treated by means of the 
variational formalism developed by Schutz \cite{schutz}. The appropriate 
Wheeler-DeWitt equation for the scale factor has the form of a Schr\"{o}dinger 
equation for a potential with a barrier. We solved it, numerically, and determined
the tunneling probability for the birth of an asymptotically DeSitter, inflationary 
universe, as a function of the radiation energy and the cosmological constant.
In particular, our treatment of the problem is more general than previous ones, 
because we took into account the fact that the scale factor ($a$) cannot be smaller 
than zero. It means that, one has to introduce an infinity potential wall at $a = 0$, 
which forces any wave-packet to be zero there. As we saw, that condition introduced 
new results, in comparison with previous works. Then, we verified that the tunneling probability increases with the radiation energy, for a fixed cosmological constant. 
Therefore, it is more probable that the classical evolution should start with the 
greatest possible value for the radiation energy. We also gave a detailed comparison 
between the exact TP, computed here, and the corresponding WKB tunneling probability. 
Finally, we evaluated the dependence of the exact TP with the cosmological constant. 
We obtained that, the TP increases with the cosmological constant for a fixed radiation 
energy. Therefore, it is more probable that the classical evolution should start with 
the greatest possible value for the cosmological constant. Also, if one assumes that 
the cosmological constant describes a vacuum energy, the latter result seems to be in 
accordance with the predictions of the {\it tunneling} wave function \cite{vilenkin}.

\begin{acknowledgements}
G. Oliveira-Neto thanks F. Takakura for the opportunity to use the Computer 
Laboratory of the Departament of Physics, UFJF, where part of this work was 
prepared. E. V. Corr\^{e}a Silva (Researcher of CNPq, Brazil), L. G. Ferreira 
Filho, G. A. Monerat and G. Oliveira-Neto thanks CNPq for partial financial 
support (Edital Universal CNPq/476852/2006-4). G. A. Monerat thanks FAPERJ for 
partial financial support (contract E-26/170.762/2004). P. Romildo Jr. thanks
CAPES of Brazil, for his scholarship.
\end{acknowledgements}

\appendix

\section{Tables}

\begin{table}[h!]
{\scriptsize\begin{tabular}{|c|c|c|c|c|c|}
\hline Energia & $TP_{int}$ & $TP_{WKB}$ & $\tau$ & $altp$ & $artp$ \\ \hline 
0.0000 & 0.0000 & $7.0246\times 10^{-522}$ & $\infty$ & 0.0000 & 17.3205 \\ \hline 
1.0000 & $2.5795\times 10^{-67}$ & $2.7574\times 10^{-517}$ & $4.4790\times 10^{+66}$ 
& 0.5777 & 17.3109 \\ \hline 
2.0000 & $3.9975\times 10^{-64}$ & $2.7181\times 10^{-513}$ & $4.0896\times 10^{+63}$ 
& 0.8174 & 17.3012 \\ \hline 
3.0000 & $4.8040\times 10^{-61}$ & $1.5939\times 10^{-509}$ & $4.1702\times 10^{+60}$ 
& 1.0017 & 17.2915 \\ \hline 
4.0000 & $2.6388\times 10^{-59}$ & $6.6774\times 10^{-506}$ & $8.7714\times 10^{+58}$ 
& 1.1573 & 17.2818 \\ \hline 
5.0000 & $5.7738\times 10^{-57}$ & $2.1799\times 10^{-502}$ & $4.4844\times 10^{+56}$ 
& 1.2946 & 17.2721 \\ \hline 
6.0000 & $9.3459\times 10^{-56}$ & $5.8369\times 10^{-499}$ & $3.0366\times 10^{+55}$ 
& 1.4190 & 17.2623 \\ \hline 
7.0000 & $6.9178\times 10^{-55}$ & $1.3258\times 10^{-495}$ & $4.4337\times 10^{+54}$ 
& 1.5335 & 17.2525 \\ \hline 
8.0000 & $7.0061\times 10^{-56}$ & $2.6169\times 10^{-492}$ & $4.6827\times 10^{+55}$ 
& 1.6404 & 17.2427 \\ \hline 
9.0000 & $6.3878\times 10^{-53}$ & $4.5691\times 10^{-489}$ & $5.4506\times 10^{+52}$ 
& 1.7409 & 17.2328 \\ \hline 
10.0000 & $3.9939\times 10^{-51}$ & $7.1563\times 10^{-486}$ & $9.1944\times 10^{+50}$ 
& 1.8361 & 17.2229 \\ \hline 
15.0000 & $1.3310\times 10^{-46}$ & $1.8319\times 10^{-470}$ & $3.3888\times 10^{+46}$ 
& 2.2553 & 17.1731 \\ \hline 
20.0000 & $3.3918\times 10^{-44}$ & $9.0816\times 10^{-456}$ & $1.5401\times 10^{+44}$ 
& 2.6119 & 17.1224 \\ \hline 
30.0000 & $1.4814\times 10^{-41}$ & $7.5933\times 10^{-428}$ & $4.3450\times 10^{+41}$ 
& 3.2183 & 17.0189 \\ \hline 
40.0000 & $5.8991\times 10^{-40}$ & $2.5466\times 10^{-401}$ & $1.2679\times 10^{+40}$ 
& 3.7397 & 16.9120 \\ \hline 
50.0000 & $9.8017\times 10^{-39}$ & $8.0358\times 10^{-376}$ & $8.5875\times 10^{+38}$ 
& 4.2086 & 16.8014 \\ \hline 
60.0000 & $1.1252\times 10^{-37}$ & $3.9314\times 10^{-351}$ & $8.2507\times 10^{+37}$ 
& 4.6419 & 16.6869 \\ \hline 
70.0000 & $1.1121\times 10^{-36}$ & $4.1409\times 10^{-327}$ & $9.0821\times 10^{+36}$ 
& 5.0499 & 16.5680 \\ \hline 
80.0000 & $1.0627\times 10^{-35}$ & $1.1845\times 10^{-303}$ & $1.0236\times 10^{+36}$ 
& 5.4391 & 16.4443 \\ \hline 
90.0000 & $1.0557\times 10^{-34}$ & $1.0939\times 10^{-280}$ & $1.1016\times 10^{+35}$ 
& 5.8147 & 16.3153 \\ \hline 
100.0000 & $1.1488\times 10^{-33}$ & $3.7271\times 10^{-258}$ & $1.0760\times 10^{+34}$ 
& 6.1803 & 16.1803 \\ \hline 
110.0000 & $1.4319\times 10^{-32}$ & $5.2113\times 10^{-236}$ & $9.1338\times 10^{+32}$ 
& 6.5393 & 16.0386 \\ \hline 
120.0000 & $2.1333\times 10^{-31}$ & $3.2602\times 10^{-213}$ & $6.4634\times 10^{+31}$ 
& 6.8942 & 15.8893 \\ \hline 
130.0000 & $3.9754\times 10^{-30}$ & $9.8051\times 10^{-193}$ & $3.6464\times 10^{+30}$ 
& 7.2479 & 15.7311 \\ \hline 
140.0000 & $9.7584\times 10^{-29}$ & $1.5060\times 10^{-171}$ & $1.5582\times 10^{+29}$ 
& 7.6029 & 15.5626 \\ \hline 
150.0000 & $3.3597\times 10^{-27}$ & $1.2439\times 10^{-150}$ & $4.7398\times 10^{+27}$ 
& 7.9623 & 15.3819 \\ \hline 
160.0000 & $1.7562\times 10^{-25}$ & $5.7774\times 10^{-130}$ & $9.4858\times 10^{+25}$ 
& 8.3293 & 15.1863 \\ \hline 
165.0000 & $1.5321\times 10^{-24}$ & $1.0157\times 10^{-119}$ & $1.1118\times 10^{+25}$ 
& 8.5171 & 15.0818 \\ \hline 
170.0000 & $1.5472\times 10^{-23}$ & $1.5685\times 10^{-109}$ & $1.1259\times 10^{+24}$ 
& 8.7085 & 14.9720 \\ \hline 
175.0000 & $1.8431\times 10^{-22}$ & $2.1361\times 10^{-99}$ & $9.6624\times 10^{+22}$ 
& 8.9045 & 14.8563 \\ \hline 
180.0000 & $2.6520\times 10^{-21}$ & $2.5758\times 10^{-89}$ & $6.8673\times 10^{+21}$ 
& 9.1059 & 14.7337 \\ \hline 
185.0000 & $4.7418\times 10^{-20}$ & $2.7600\times 10^{-79}$ & $3.9286\times 10^{+20}$ 
& 9.3142 & 14.6029 \\ \hline 
190.0000 & $1.0919\times 10^{-18}$ & $2.6372\times 10^{-69}$ & $1.7457\times 10^{+19}$ 
& 9.5310 & 14.4624 \\ \hline 
195.0000 & $3.3916\times 10^{-17}$ & $2.2544\times 10^{-59}$ & $5.7545\times 10^{+17}$ 
& 9.7585 & 14.3099 \\ \hline 
200.0000 & $1.5114\times 10^{-15}$ & $1.7295\times 10^{-49}$ & $1.3233\times 10^{+16}$ 
& 10.0000 & 14.1421 \\ \hline 
205.0000 & $1.0542\times 10^{-13}$ & $1.1943\times 10^{-39}$ & $1.9466\times 10^{+14}$ 
& 10.2605 & 13.9543 \\ \hline 
210.0000 & $1.3129\times 10^{-11}$ & $7.4432\times 10^{-30}$ & $1.6069\times 10^{+12}$ 
& 10.5485 & 13.7379 \\ \hline 
215.0000 & $3.6494\times 10^{-09}$ & $4.1983\times 10^{-20}$ & $5.9628\times 10^{+09}$ 
& 10.8801 & 13.4767 \\ \hline 
216.0000 & $1.2796\times 10^{-08}$ & $3.7003\times 10^{-18}$ & $1.7121\times 10^{+09}$ 
& 10.9545 & 13.4164 \\ \hline 
217.0000 & $4.7368\times 10^{-08}$ & $3.2487\times 10^{-16}$ & $4.6582\times 10^{+08}$ 
& 11.0325 & 13.3523 \\ \hline 
218.0000 & $1.8642\times 10^{-07}$ & $2.8413\times 10^{-14}$ & $1.1926\times 10^{+08}$ 
& 11.1150 & 13.2837 \\ \hline 
219.0000 & $7.8683\times 10^{-07}$ & $2.4754\times 10^{-12}$ & $2.8476\times 10^{+07}$ 
& 11.2029 & 13.2097 \\ \hline 
220.0000 & $3.6052\times 10^{-06}$ & $2.1485\times 10^{-10}$ & $6.2674\times 10^{+06}$ 
& 11.2978 & 13.1286 \\ \hline 
221.0000 & $1.8228\times 10^{-05}$ & $1.8577\times 10^{-8}$ & $1.2512\times 10^{+06}$ 
& 11.4018 & 13.0384 \\ \hline 
222.0000 & $1.0419\times 10^{-04}$ & $1.6002\times 10^{-6}$ & $2.2110\times 10^{+05}$ 
& 11.5187 & 12.9352 \\ \hline 
223.0000 & $7.0045\times 10^{-04}$ & $1.3731\times 10^{-4}$ & $3.3281\times 10^{+04}$ 
& 11.6558 & 12.8118 \\ \hline 
224.0000 & $5.9816\times 10^{-03}$ & $1.1671\times 10^{-2}$ & $3.9562\times 10^{+03}$ 
& 11.8322 & 12.6491 \\ \hline 
\end{tabular}
}
\caption{{\protect\footnotesize {The computed values of $TP_{int}$,
$TP_{WKB}$, $\tau$, $a_{ltp}$ and $a_{rtp}$ for $47$ different values of $E$
when $\Lambda=0.01$.}}}
\label{tableenergy}
\end{table}

\begin{table}[h!]
{\scriptsize\begin{tabular}{|c|c|c|c|c|c|}
\hline $\Lambda$ & $TP_{int}$ & $PE_{max}$ & $\tau$ & $a_{ltp}$ & $a_{rtp}$\\ \hline
$0.0100$ & $4.7449\times 10^{-20}$ & $2.2500\times 10^{+02}$ & $3.9260\times 10^{+20}$ &
$9.3142$ & $14.6029$\\ \hline
$0.0101$ & $1.9386\times 10^{-17}$ & $2.2277\times 10^{+02}$ & $9.6427\times 10^{+17}$ &
$9.3467$ & $14.4800$\\ \hline
$0.0102$ & $5.5162\times 10^{-15}$ & $2.2059\times 10^{+02}$ & $3.4010\times 10^{+15}$ &
$9.3803$ & $14.3571$\\ \hline
$0.0103$ & $1.0768\times 10^{-12}$ & $2.1845\times 10^{+02}$ & $1.7487\times 10^{+13}$ &
$9.4152$ & $14.2343$\\ \hline
$0.0104$ & $1.4239\times 10^{-10}$ & $2.1635\times 10^{+02}$ & $1.3276\times 10^{+11}$ &
$9.4515$ & $14.1114$\\ \hline
$0.0105$ & $1.2522\times 10^{-08}$ & $2.1429\times 10^{+02}$ & $1.5156\times 10^{+09}$ &
$9.4892$ & $13.9882$\\ \hline
$0.0106$ & $7.1354\times 10^{-07}$ & $2.1226\times 10^{+02}$ & $2.6708\times 10^{+07}$ &
$9.5286$ & $13.8645$\\ \hline
$0.0107$ & $2.5363\times 10^{-05}$ & $2.1028\times 10^{+02}$ & $7.5462\times 10^{+05}$ &
$9.5697$ & $13.7402$\\ \hline
$0.0108$ & $5.3391\times 10^{-04}$ & $2.0833\times 10^{+02}$ & $3.6009\times 10^{+04}$ &
$9.6129$ & $13.6151$\\ \hline
$0.0109$ & $6.1795\times 10^{-03}$ & $2.0642\times 10^{+02}$ & $3.1259\times 10^{+03}$ &
$9.6583$ & $13.4888$\\ \hline
$0.0110$ & $3.5077\times 10^{-02}$ & $2.0455\times 10^{+02}$ & $5.5342\times 10^{+02}$ &
$9.7062$ & $13.3610$\\ \hline
$0.0111$ & $8.4175\times 10^{-02}$ & $2.0270\times 10^{+02}$ & $2.3183\times 10^{+02}$ &
$9.7570$ & $13.2314$\\ \hline
$0.0112$ & $9.5984\times 10^{-02}$ & $2.0089\times 10^{+02}$ & $2.0443\times 10^{+02}$ &
$9.8112$ & $13.0996$\\ \hline
$0.0113$ & $1.2079\times 10^{-01}$ & $1.9912\times 10^{+02}$ & $1.6341\times 10^{+02}$ &
$9.8692$ & $12.9648$\\ \hline
$0.0114$ & $1.3117\times 10^{-01}$ & $1.9737\times 10^{+02}$ & $1.5143\times 10^{+02}$ &
$9.9318$ & $12.8264$\\ \hline
$0.0115$ & $1.4639\times 10^{-01}$ & $1.9565\times 10^{+02}$ & $1.3662\times 10^{+02}$ &
$10.0000$ & $12.6834$\\ \hline
$0.0116$ & $1.6190\times 10^{-01}$ & $1.9397\times 10^{+02}$ & $1.2446\times 10^{+02}$ &
$10.0752$ & $12.5344$\\ \hline
$0.0117$ & $1.7538\times 10^{-01}$ & $1.9231\times 10^{+02}$ & $1.1586\times 10^{+02}$ &
$10.1594$ & $12.3773$\\ \hline
$0.0118$ & $1.8752\times 10^{-01}$ & $1.9068\times 10^{+02}$ & $1.0938\times 10^{+02}$ &
$10.2559$ & $12.2088$\\ \hline
$0.0119$ & $1.9940\times 10^{-01}$ & $1.8908\times 10^{+02}$ & $1.0402\times 10^{+02}$ &
$10.3703$ & $12.0232$\\ \hline
$0.0120$ & $2.1463\times 10^{-01}$ & $1.8750\times 10^{+02}$ & $9.7983\times 10^{+01}$ &
$10.5150$ & $11.8082$\\ \hline
$0.0121$ & $2.2964\times 10^{-01}$ & $1.8595\times 10^{+02}$ & $9.3439\times 10^{+01}$ &
$10.7287$ & $11.5252$\\ \hline
\end{tabular}
}
\caption{{\protect\footnotesize {The computed values of $TP_{int}$, $PE_{max}$,
$\tau$, $a_{ltp}$ and $a_{rtp}$ for $22$ different values of $\Lambda$
when $E=185$.}}}
\label{tablelambda}
\end{table}


\begin{thebibliography}{99}

\bibitem{dewitt} B. S. DeWitt, Phys. Rev. D \textbf{160}, 1113 (1967).

\bibitem{grishchuk} L. P. Grishchuk and Ya. B. Zeldovich, in {\it Quantum
Structure of Space and Time}, eds. M. Duff and C. Isham (Cambridge University
Press, Cambridge, 1982).

\bibitem{vilenkin} A. Vilenkin, Phys. Lett. B \textbf{117}, 25 (1982); Phys.
Rev. D \textbf{30}, 509 (1984); ibid. \textbf{33}, 3560 (1986).

\bibitem{hawking} J. B. Hartle and S. W. Hawking, Phys. Rev. D \textbf{28},
2960 (1983).

\bibitem{linde} A. D. Linde, Lett. Nuovo Cim. \textbf{39}, 401 (1984).

\bibitem{rubakov} V. A. Rubakov, Phys. Lett. B \textbf{148}, 280 (1984).

\bibitem{vilenkin1} For a recent critical review see: A. Vilenkin, in
{\it Cambridge 2002, The future of theoretical physics and cosmology}, 
eds. G. W. Gibbons, E. P. S. Shellard and S. J. Rankin (Cambridge University
Press, Cambridge, 2003), 649-666.

\bibitem{wheeler} J. A. Wheeler, in {\it Batelles Rencontres}, eds. C. DeWitt
and J. A. Wheeler (Benjamin, New York, 1968), 242.

\bibitem{hawking1} S. W. Hawking and N. Turok, Phys. Lett B \textbf{425}, 25 
(1998).

\bibitem{linde1} A. Linde, Phys. Rev. D \textbf{58}, 083514 (1998); ibid.
\textbf{59}, 023503 (1998).

\bibitem{bousso} R. Bousso and A. Linde, Phys. Rev. D \textbf{58}, 083503
(1998).

\bibitem{rubakov1} D. Levkov, C. Rebbi and V. A. Rubakov, Phys. Rev. D 
\textbf{66}, 083516 (2002).

\bibitem{vilenkin2} A. Vilenkin, Phys. Rev. D \textbf{37}, 888 (1988);
T. Vachaspati and A. Vilenkin, ibid. \textbf{37}, 898 (1988); J. Garriga and
A. Vilenkin, ibid. \textbf{56}, 2464 (1997); J. Hong, A. Vilenkin and S.
Winitzki, ibid. \textbf{68}, 023520 (2003); \textbf{68}, 023521 (2003).

\bibitem{schutz} Schutz, B. F., Phys. Rev. D \textbf{2}, 2762, (1970);
Schutz, B. F., Phys. Rev. D \textbf{4}, 3359, (1971).

\bibitem{bertolami} O. Bertolami and J. M. Mour\~{a}o, Class. Quantum Grav. 
\textbf{8}, 1271 (1991).

\bibitem{cavaglia} M. Cavaglia, V. Alfaro and A. T. Filippov, Int. J. Mod.
Phys. A \textbf{10}, 611 (1995).

\bibitem{gil} Y. Fujiwara et al., Class. Quantum Grav. \textbf{7} 163
(1992); Phys. Rev. D \textbf{44}, 1756 (1991); J. Louko and P. J. Ruback,
Class. Quantum Grav. \textbf{8}, 91 (1991); J. J. Halliwell and J. Louko,
Phys. Rev. D {\b 42}, 3997 (1990); G. Oliveira-Neto, Phys. Rev. D \textbf{58}, 
107501 (1998).

\bibitem{moss} Mariam Bouhmadi-Lopez and Paulo Vargas Moniz, Phys. Rev. D 
\textbf{71}, 063521 (2005); I. G. Moss and W. A. Wright, Phys. Rev. D 
\textbf{29}, 1067 (1984); M. J. Gotay and J. Demaret, Phys. Rev. D \textbf{28
}, 2402 (1983); G. A. Monerat, E. V. Corr\^{e}a Silva, G. Oliveira-Neto,
L. G. Ferreira Filho and N. A. Lemos, Phys. Rev. D \textbf{73}, 044022 (2006). 

\bibitem{germano1} F. G. Alvarenga, J. C. Fabris, N. A. Lemos, G. A. 
Monerat, Gen. Rel. Grav. {\bf 34}, 651 (2002).

\bibitem{abramowitz} M. Abramowitz and I. Stegun, \textit{Handbook of
Mathematical Functions}, (Dover Publications Inc., New York, 1965), p. 1046.

\bibitem{dirac} P. A. M. Dirac, Can. J. Math. \textbf{2}, 129 (1950); Proc.
Roy. Soc. London A \textbf{249}, 326 and 333 (1958); Phys. Rev. \textbf{114},
924 (1959).

\bibitem{lemos} N. A. Lemos, J. Math. Phys. \textbf{37}, 1449 (1996).

\bibitem{crank} J. Crank and P. Nicholson, Proc. Cambridge Philos. Soc.
\textbf{43}, 50 (1947).

\bibitem{numericalrecipes} For a more detailed explanation see: W. H. Press,
S. A. Teukolsky, W. T. Vetterling and B. P. Flannery, \textit{Numerical 
Recipes}, (Cambridge University Press, Cambridge, England, 1997), Sec. 19.2; 
and C. Scherer, \textit{M\'{e}todos Computacionais da F\'{\i}sica}, (Editora 
Livraria da F\'{\i}sica, S\~{a}o Paulo, 2005), Chap. 3.

\bibitem{Iitaka} T. Iitaka, Phys. Rev. E \textbf{49}, 4684 (1994).

\bibitem{teukolsky} S. A. Teukolsky, Phys. Rev. D \textbf{61}, 087501 (2000).

\bibitem{griffths} D. J. Griffths, \textit{Introduction to Quantum Mechanics},
(Prentice-Hall, Inc., Englewood Cliffs, 1995), Chap. 8; and L. D. Landau and 
E. M. Lifshitz, \textit{Quantum Mechanics: Non-Relativistic Theory}, (Pergamon 
Press, London, 1959), Chap. III.

\bibitem{merzbacher} E. Merzbacher, \textit{Quantum Mechanics}, (John Wiley \& 
Sons, Inc., New York, 1998, Ed. 3), Chap. 7.

\end{thebibliography}
\end{document}